\documentclass[12pt,preprint]{aastex}
\usepackage{multirow}
\pdfoutput=1

\begin{document}

\newcommand{\gsim}{\raisebox{-.4ex}{$\stackrel{>}{\scriptstyle \sim}$}}
\newcommand{\lsim}{\raisebox{-.4ex}{$\stackrel{<}{\scriptstyle \sim}$}}
\newcommand{\psim}{\raisebox{-.4ex}{$\stackrel{\propto}{\scriptstyle \sim}$}}
\newcommand{\kms}{\mbox{km~s$^{-1}$}}
\newcommand{\s}{\mbox{$''$}}
\newcommand{\mloss}{\mbox{$\dot{M}$}}
\newcommand{\my}{\mbox{$M_{\odot}$~yr$^{-1}$}}
\newcommand{\ls}{\mbox{$L_{\odot}$}}
\newcommand{\ms}{\mbox{$M_{\odot}$}}
\newcommand{\mm}{\mbox{$\mu$m}}
\newcommand{\water}{\mbox{H$_2$O}}
\newcommand{\methanol}{\mbox{CH$_3$OH}}
\def\arcdeg{\hbox{$^\circ$}}
\newcommand{\secp}{\mbox{\rlap{.}$''$}}%

\newcommand{\etal}{{et al.\ }}
\newcommand{\mv}{{MultiView }}
\newcommand{\cone}{{J0121+1149 }} 
\newcommand{\ctwo}{{J0113+1324 }} 
\newcommand{\cthree}{{J0042+1009 }} 
\newcommand{\cfour}{{J0106+1300 }} 
\newcommand{\ohm}{{OH128.6-50.1 }} 

\shortauthors{Rioja, et al.}
\shorttitle{MultiView High Precision VLBI Astrometry at Low Frequencies} 

\title{ MultiView High Precision VLBI Astrometry at Low Frequencies}


 \author{
   Mar\'{\i}a J. \textsc{Rioja}\altaffilmark{1,2,3},
   Richard \textsc{Dodson}\altaffilmark{1},
   Gabor \textsc{Orosz}\altaffilmark{4},
   Hiroshi \textsc{Imai}\altaffilmark{4,5},
   Sandor \textsc{Frey}\altaffilmark{6}
}
\affil{$^1$ ICRAR, M468, University of Western Australia, 35 Stirling
  Hwy, Perth 6009, Australia} 
\affil{$^2$ CSIRO Astronomy and Space Science, 26 Dick Perry Avenue, Kensington WA 6151, Australia}   
\affil{$^3$ OAN (IGN), Alfonso XII, 3 y 5, 28014 Madrid, Spain}
\affil{$^4$ Graduate School of Science and Engineering, Kagoshima University, 1-21-35 Korimoto, Kagoshima 890-0065, Japan}
\affil{$^5$ Science and Engineering Area of Research and Education
  Assembly, Kagoshima University, 1-21-35 Korimoto, Kagoshima 890-0065, Japan}
\affil{$^6$ Konkoly Observatory, MTA Research Centre for Astronomy and Earth Sciences, PO Box 67, H-1525 Budapest, Hungary}
 \email{maria.rioja@icrar.org}

\begin{abstract} 

The arrival of the Square Kilometer Array (SKA) will revitalise all aspects
of Very Long Baseline Interferometry  (VLBI) astronomy at the lower
frequencies. In the last decade there have
been huge strides towards routinely achieving high precision VLBI astrometry at
frequencies dominated by the tropospheric contributions, most notably
at 22GHz, using advanced phase referencing techniques. Nevertheless
to increase the capability for high precision astrometric measurements
at low radio frequencies ($<$8~GHz) an effective calibration strategy of the
systematic ionospheric propagation effects that is widely applicable
is required.
Observations at low frequencies are dominated by
distinct direction dependent ionospheric propagation errors, which
place a very tight limit on the angular separation of a suitable
phase referencing calibrator.
 
The MultiView technique holds the key to the compensation of
atmospheric spatial-structure errors,
by using observations of multiple calibrators and 2-D interpolation. 
In this paper we present the first demonstration of the power of
MultiView using three calibrators, several degrees from the target, 
along with a  comparative study of the astrometric accuracy between MultiView and
phase-referencing techniques.


MultiView calibration provides an order of magnitude
improvement in astrometry with respect to conventional phase
referencing, achieving $\sim100$micro-arcseconds astrometry errors in
a single epoch of observations, effectively reaching the thermal noise
limit.

MultiView will achieve its full
potential with the enhanced sensitivity and multibeam capabilities of
SKA and the pathfinders, which will enable simultaneous observations
of the target and calibrators. Our demonstration indicates that the
10~micro-arcseconds goal of astrometry at ~1.6GHz using VLBI with SKA
is feasible using the MultiView technique.

\end{abstract}

\section{Introduction}\label{sec:intr}

Very Long Baseline Interferometry (VLBI) observations hold the potential to achieve the highest astrometric accuracy in
astronomy, provided that the fringe phase observable can be calibrated
\citep{alef_88}. 
The development of advanced phase referencing (PR) techniques to compensate
for the tropospheric propagation errors have led to routinely
achieving micro-arcsecond ($\mu$as) astrometry 
at frequencies between ca. 10 and a few tens of GHz 
using alternating observations of the target and a nearby
calibrator, which can be up to a few degrees away
\citep[]{reid_04,honma_08_trop}. 
The increasingly fast tropospheric fluctuations at higher
frequencies set an upper threshold for application of PR techniques at
$\sim \,$43 GHz (with but a single case at 86 GHz \citep{porcas_02}).
More recently, the development of phase calibration techniques using (nearly) simultaneous
observations at multiple mm-wavelengths, that
is Source Frequency Phase Referencing (SFPR) \citep{rioja_11a} and
Multi Frequency Phase Referencing (MFPR) \citep{dodson_16}, 
have extended the
capability to measure  $\mu$as astrometry up to mm-wavelengths. 
This capability for accurate astrometry has resulted in a wide applicability to many
scientific problems \citep[and references therein]{reid_micro}. 
Nevertheless the application of these advanced PR techniques 
to relatively low frequencies $\le8$~GHz are 
hindered by the contribution of ionospheric propagation effects,
increasingly dominant at lower frequencies, which have a different
nature to the tropospheric effects.
The unpredictability of
the spatial irregularities in the plasma density in the ionosphere
introduces differential path variations between the sky directions of
the two sources, and propagate into systematic position errors even
for small source separations. 
In addition, at the lowest frequencies,
the temporal variations are also an issue.
These both are responsible for  
degrading the positional accuracy achieved with this technique and,
eventually, prevent the phase connection process and the use of conventional phase referencing.

Therefore, a new strategy is required to overcome the limitations
imposed by the ionospheric propagation medium and to reach the full potential
of the instruments working on these spectral regimes, such as the
Square Kilometer Array (SKA)
that will have VLBI capability between 0.3 and 14~GHz.
In general, observations which involve more than one calibrator have demonstrated advantages
for astrometric VLBI at low frequencies \citep{fomalont_02, rioja_02, doi_06}. The alternative is the unusual configuration
when a target and a strong calibrator lie within the field-of-view
(FoV) of the VLBI antennas
(i.e. an `in-beam' calibrator), and thus can be observed simultaneously
\citep{vlba_24,fomalont_99}. A useful variation of this combines
the observations of a weak `in-beam'  calibrator source and nodding to
a strong more distant calibrator \citep{doi_06}. The observations of the strong calibrator are used to remove the first-order atmospheric
effects; then the observations of the weak `in-beam' source, which is observed along with the target source,
provide further adjustments of the spatial and
temporal fluctuations, with longer coherence times. 
The results obtained with this approach are positive, however its widespread
application is still limited by sensitivity. Another useful approach is when there are two calibrators
aligned with, but on opposite sides of, the target \citep{fomalont_02}. During the observations the telescopes alternate
every few minutes between the three sources, and in the analysis successive scans on the calibrators
are used for the spatial and temporal interpolation to the enclosed position and scan time of the
target source. The rare source configuration required for this approach to work results in
limited applicability, and the calibration time overhead is large.

In this paper we present results from the MultiView technique 
which, by deriving 2D phase screens 
from observations of three or more calibrators,
achieves a superior mitigation of atmospheric errors that results in
increased precision astrometry,  along with wide applicability by relaxing the constraints on the angular
separation up to few degrees, and does not require alignment of
sources. The scope of application is for the low frequency regime where
the perfomance of PR is degraded due to the spatial structure of
the ionospheric dominant errors.
It is a development of the ``cluster--cluster'' VLBI technique, which
allowed simultaneous observations of a target and multiple calibrators around it by replacing single
telescopes by sites with multiple elements \citep{rioja_97}. The ability of the ``cluster--cluster'' technique to address the ionospheric
effects has been demonstrated with joint observations between connected interferometer
arrays at 1.6 GHz of a target and three calibrator sources
\citep{rioja_02, rioja_09}. Despite these benefits its
use has been limited by the shortage of compatible observing sites, and the complexity in its implementation.

We revisit this technique in the light of the next generation of instruments for low
frequency observations that will become operational in the course of
the next decade.
These have the multi-beam capability as an “in-built” feature, such as
the Australian SKA Pathfinder (ASKAP) in the near future, and SKA
in the longer term. We believe that the implementation of MultiView techniques will enhance
the performance of VLBI observations, by providing higher precision astrometric
measurements of many targets
at low frequencies.

In this paper we describe the observations in Section~\ref{sec:obs}; the basis of the MultiView technique in Section 
~\ref{sec:meth}; Section~\ref{sec:res} presents
a demonstration and quantifies its astrometric capabilities, along with a comparative study with outcomes from
conventional PR techniques for a range of target--calibrator angular separations, including in-beam phase
referencing, using VLBA observations at 1.6 GHz; Section~\ref{sec:disc} are discussions and conclusions.

\section{Observations}\label{sec:obs}

\begin{figure}[htbp]
\centering
\includegraphics[width=0.9\textwidth]{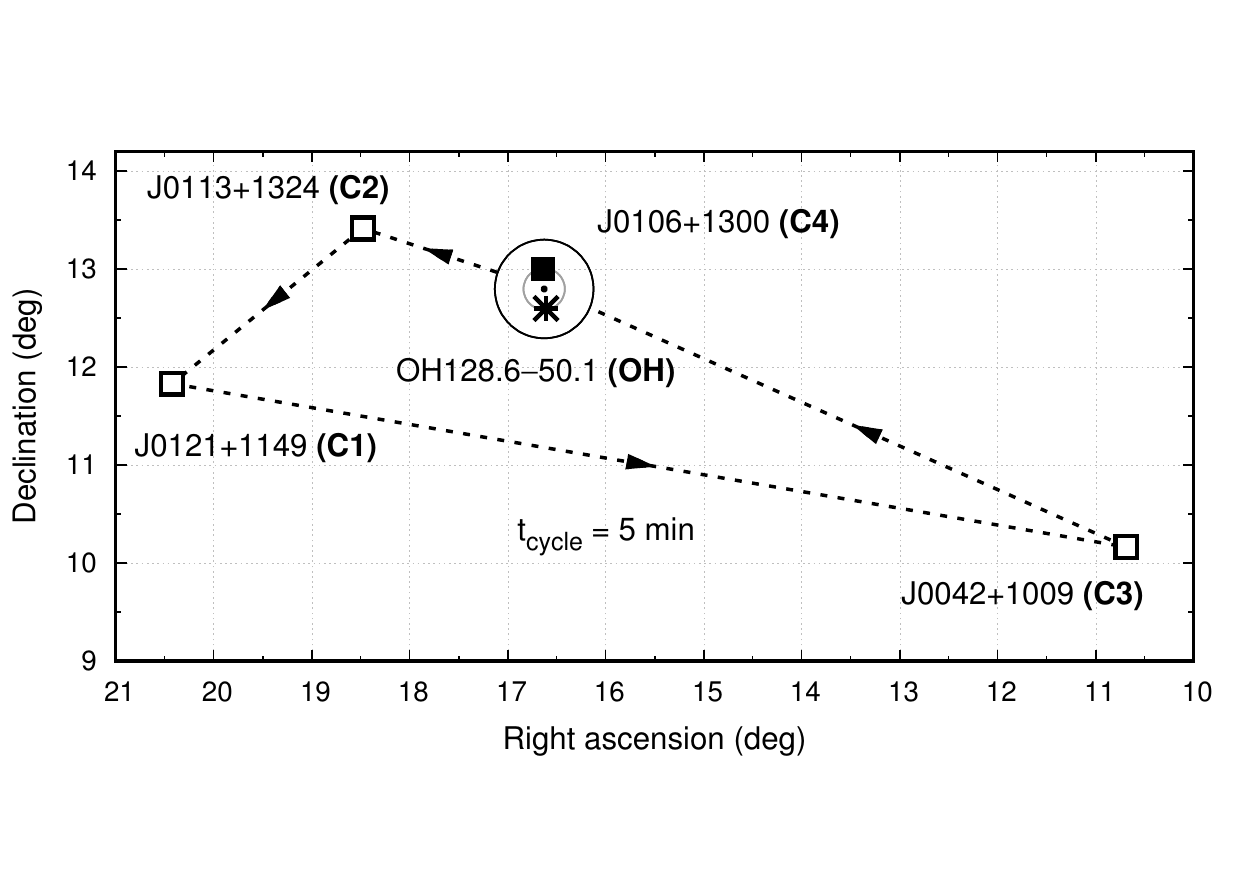}
\caption{\label{fig:obs_setup}Sky distribution of the sources observed
with the VLBA at 1.6 GHz. Table~\ref{table:sources} lists the source coordinates. Dashed lines and arrows mark the source switching order
  during the observations with 5-min duty cycles. Star and solid
  symbols mark the simultaneously observed OH--C4 pair, with the
  VLBA antennas pointed halfway between the two. The two concentric
  circles represent the half-power beamwidth  and full beamwidth of the antennas. Both OH and
  C4 are targets in the astrometric analyses (see
  Table~\ref{tab:analyses} for description of analyses). C1 was used as the fringe-finder.}
\end{figure}

In an effort to demonstrate the improvements by MultiView calibration
over conventional techniques, we conducted two epochs of observations
with the NRAO\footnote{The National Radio Astronomy Observatory is a
  facility of the National Science Foundation operated under
  cooperative agreement by Associated Universities, Inc.} Very Long
Baseline Array (VLBA) separated by one month, on 2015 June 8 (Epoch I,
obs. id:~BO047A7) and July 7 (Epoch II, obs. id:~BO047A4),
at 1.6 GHz.  Both epochs of observations used
identical setups with a duration of $\sim\, 4$
hours. Table~\ref{table:sources} lists the source names and
coordinates and Fig.~\ref{fig:obs_setup} shows the distribution in the
sky.

\begin{table}[hbtp]
\centering
\begin{tabular}{cclllcc}
\hline
\hline
Source Name & Alias & \multicolumn{1}{c}{Right Ascension} &
                                                            \multicolumn{1}{c}{Declination}
  & $\sigma_{RA}$ & $\sigma_{DEC}$ & S$_{\rm 1.6~GHz}$ \\
&& \multicolumn{1}{c}{( h \enspace m \enspace s )} &
                                                     \multicolumn{1}{c}{(
                                                     \degr \enspace
                                                     \arcmin \enspace
                                                     \arcsec )} &
                                                                  (mas)
  & (mas) & (Jy beam$^{-1}$) \\
\hline
J0121+1149 & C1     & 01 21 41.595044 & +11 49 50.41304 & 0.10 & 0.10
                                   & 2.1 \\
J0113+1324  & C2   & 01 13 54.510365 & +13 24 52.47783 & 0.26 & 0.38 &
                                                                       0.08 \\
J0042+1009   & C3  & 00 42 44.371738 & +10 09 49.20750 & 0.15 & 0.17 &
                                                                       0.17\\
J0106+1300    & C4 & 01 06 33.356509 & +13 00 02.60390 & 0.14 &
                                                                0.19&0.07 \\ \hline
OH128.6--50.1 & OH &01 06 25.98     & +12 35 53.0     &      &      &
                                                                      0.11 \\
\hline
\end{tabular}
\caption{All observed sources: four quasars and one OH-maser line
  source (also known as WX Psc).
  Columns 1 and 2 are the source names and aliases used through this
paper, respectively. Columns 3 and 4 are the Right Ascension  and
Declination coordinates used at the correlator. Columns 5 and 6
are the corresponding position errors, if available.  Column 7 is the
peak brightness at 1.6~GHz measured from our observations. For quasars, all quantities are from the VLBA
Radio Fundamental Catalog 
(L. Petrov, solution rfc\_2015b (unpublished) available on the Web at http://asrtogeo.org/vlbi/solutions/rfc\_2015b).
For the OH-maser, position comes  from the SIMBAD Astronomical
Database (http://adsabs.harvard.edu/abs/2000A\%26AS..143....9W)}
\label{table:sources}
\end{table}

The observations consisted of alternated scans switching between all the sources with a duty cycle of $\sim\,5$ minutes.
The two sources in the centre of the distribution, the OH
maser source and the quasar C4, were observed simultaneously 
because they lie within the primary beam of the VLBA antennas. They are the targets of the analyses presented in this paper,
allowing the MultiView calibration to be tested for both a maser line and quasar
continuum observations simultaneously. 
The sessions were long enough to ensure
sufficient $(u,v)$ coverage and sensitivity, spending $\sim$40\% of the
time on the OH--C4 pair with alternating 30--90 s scans on the C1, C2
and C3 calibrators. Both epochs were observed at a
similar time range around early morning, 12:06:00$-$16:06:00 UT for Epoch I
and 11:12:00$-$15:12:00 UT for Epoch II, when the variations in the
ionosphere and its effects on astrometry are expected to be the
largest.

The 2-bit quantized signals were recorded in dual circular
polarization with 256 Mbps using 4 intermediate frequency (IF) bands,
each with a bandwidth of 8~MHz. The IFs were spread out over 300~MHz,
centered around the four ground-state OH maser lines of 1612, 1665,
1667, 1720~MHz and the H\,{\sc I} line at 1420~MHz. Each band had a
channel spacing of 1.95~kHz, corresponding to a velocity resolution of
0.36~km~s$^{-1}$. All bands were correlated in a single run, using two
phase centers for the observations of the OH--C4 pair. 
For the OH maser source, only the 1612 MHz transition provided useful
astrometric data.

\section{Astrometric Data Analyses} 
\label{sec:meth}

We have carried out a comparative astrometric study using different 
calibration techniques and angular separations, summarized in Table~\ref{tab:analyses}, on the same observations.
The analyses comprise conventional PR techniques
using pairs of sources (i.e. one calibrator and one target
source) with angular separations of $\sim\,0.4^o$ (i.e. in-beam), $2^o$, $4^o$ and $6^o$, and MultiView
(MV) techniques using four sources (i.e. three calibrators
and one target sources). 
The targets are the quasar C$4$ and the OH maser source, while the
calibrators are quasars in all cases.
We use the alternating quasar-only observations for the comparison between
PR with a range of angular separations and MV techniques, using C4 as the target;
using the OH maser source as target, we compare in-beam PR with C4 as reference, 0.4$^o$ away, to MV using three calibrators 
$2^o$, $4^o$, and $6^o$ away. Note that
the MV observations in this paper involve source switching.
The repeatability of the astrometric measurements 
between the two epochs (i.e. inter-epoch differences) is the chosen figure of merit
for our comparative study of the calibration strategies. 
For the quasars-only analysis, the repeatabilities are  
a direct indication of the uncompensated ionospheric errors remaining after calibration,
given that no source position changes are expected. For the
OH maser analysis, the apparent source motion due to the proper motion and
parallax needs to be accounted for first, even over a 1 month timeline.
Sect.~\ref{sec:errors} 
will cover the astrometric error analysis.

The PR analyses were carried out following standard procedures in the
Astronomical Image Processing System (AIPS) software \citep{aips}.  
The MV analyses were carried out following the same procedures in AIPS
but with additional steps to incorporate direction dependent effects, as
described in Section~\ref{sec:MVbasis}. Also, we have made self calibration maps of all sources using conventional hybrid mapping
techniques in VLBI using AIPS and Difmap \citep{difmap}; these have been used to remove the source  structure
contribution, to provide reference points for the astrometric
analysis and to asses the quality of the calibration strategies using the fractional flux recovery
quantity (see Section~\ref{sec:errors}).

\subsection{Basis of MultiView Direction Dependent Phase Calibration}  \label{sec:MVbasis}

The MV calibration strategy corrects for the direction dependent 
nature of the ionospheric phase errors by using simultaneous or
near-simultaneous observations of
multiple calibrators around the target. 
Then we use a 2-D interpolation of the antenna phases, estimated along the  directions of all calibrators, to provide corrections
along the line of sight of the target observations. This is realized
by a weighted linear combination of the complex antenna gains, representing
the relative  source distribution in the sky, as shown in Fig.~\ref{fig:obs_setup} for the case of interest to this paper.
This is equivalent to the treatment of the propagation medium as 
a wedge-like spatial structure, up to several degrees in size, above
each antenna  \citep{fomalont_02,rioja_02}.
The temporal structure of the propagation medium effects is best
calibrated using simultaneous observations of the calibrators and the
target sources in MV observations. However when this observing configuration is
not possible one can use alternating observations of the sources, as
long as the duty cycle is less than the atmospheric coherence time.  

Therefore, in MV, the target position is tied to the assumed positions
of the multiple calibrators weighted as in the
analysis. That is, to
a virtual point in the sky whose location depends on the source
distribution in the sky.  Instead, in PR, the measured target
positions are tied to the assumed position of the corresponding
(single) calibrator. Nevertheless,
as long as the  calibrator sources
provide good fiducial points (i.e. are stationary), this virtual point
is also stationary and changes between the astrometric measurements at
different epochs trace the motion of the target in both MV and PR.

Our implementation of the MV direction dependent calibration strategy
is more complicated than a basic linear interpolation. It includes a
correction for untracked $2\pi$ phase ambiguities in the measured
calibrator phases, which lead to errors in the spatial interpolation.
Those are automatically detected by searching over ambiguities in the
formation of every interpolation and optimizing the result.
Additionally we have the ability to steer the corrections by
forcing the addition of ambiguities at the outset. 

\begin{table}[h]
\centering
\begin{tabular}{c c l c l }
\hline
\noalign{\smallskip}
Astrometric & Target & Reference & Angular  & Analysis \\
Calibration     & Source(s) & Source   &  Separation & Id.  \\ \hline
{\it Single Calibrator} &&&& \\ 
PR     & C4 & C2 & 1.84$^o$ & PR$_{2^o}$   \\
PR  & C4 & C1 & 3.88$^o$ & PR$_{4^o}$  \\
PR  & C4 & C3 & 6.49$^o$ &  PR$_{6^o}$  \\
`in beam' PR & OH & C4 & 0.4$^o$ & PR$_{in-beam}$  \\ \hline
{\it Three Calibrators} &&&& \\
MV        & C4 & C1, C2, C3 & $1.84^o,\, 3.88^o,\,6.49^o$ & MV$_{QSO}$  \\
MV         & OH &  C1, C2, C3 & $2.0^o,\,3.81^o,\,6.30^o$ & MV$_{OH}$
  \\ \hline
\noalign{\smallskip}
\hline
\end{tabular}
\caption{Astrometric analyses compared in
  this paper, along with aliases used
  throughout the text. Column 1 are the calibration techniques: 
  using a single calibrator, regardless that the observations of the pair are carried out using source
  switching or simultaneously observed, and
  using three calibrators; PR=Phase Referencing and MV = MultiView. 
Columns 2 and 3 list of target and  reference sources,
respectively. Colum 4 is the target--calibrator(s) angular
  separation. Column 5 is the analysis
  identification name used throughout the text. All analyses have been carried out for epochs I
and II, separated by one month.}
\label{tab:analyses} 
\end{table}

\begin{table}[h]
    \centering
\begin{tabular}{c|cc}
Target Source & Weight$_{1}$ & Weight$_{2}$ \\ \hline
\cfour (C4) &  1.147 & 0.1735  \\
\ohm   (OH) &  0.985 & 0.2475 \\
\end{tabular}
\caption{MultiView weights used 
to implement  the direction dependent ionospheric calibration along the line of
the sight of the target source in the analyses.
Column 1 lists the target sources, for analyses MV$_{QSO}$ and MV$_{OH}$, respectively.
Columns 2  and 3 are the weights that were applied for the phase transfer between
C2 and C1  (``Weight$_1$''),  and 
between C3 and this combination  (``Weight$_2$''), for the
calibration of the corresponding target source.}
    \label{table:wf}
\end{table}

\subsection{Error Analysis}\label{sec:errors}

We describe here the 
approaches used to quantify the uncertainties
of the astrometric measurements in
this paper.

1) {\it Repeatability errors:} 
We have used the repeatability of the measured positions between the two epochs of
observations, which provide independent
measurements of the relative source position, as an empirical estimate of the astrometric errors. 
The span ($\Delta_{I-II}$) of the positions are indicative of uncompensated systematic
ionospheric residual errors remaining after calibration for each analysis.
The repeatability errors,
corrected for the bias introduced by having only two measurements, are calculated as: 
$\sigma_{pos,rep} = \Delta_{I-II} / \sqrt {2} \ast \sqrt {2
    /\pi}$.
They are a measure of the precision of the calibration method, in
absence of inherent position changes, and we use them as the figure of merit for the
comparative study. Note that while there is a limited sample of two
epochs the different analyses are carried out on the same observations, enabling a direct comparison of the
compensation efficiency of the systematic errors under the same weather conditions.

2) {\it Thermal noise (and other random) errors}:
The ratio of the synthesized beam size ($\theta_B$) and the signal-to-noise ratio  (SNR) in
the astrometric maps (i.e. PRed
and MVed maps) gives an estimate of the uncertainty in the measurement of the position of a feature in
the maps due to random noise, as $\sigma_{pos,thermal} \sim 0.5
{\theta_B / SNR}$.
This error has a contribution from  thermal (usually dominated by the receiver) random noise,
and from residual atmospheric phase fluctuations. The latter depends on the
duty cycle during the observations, the angular separation between sources and the weather conditions.
It is commonly referred to as the thermal noise error and represents the ultimate astrometric
precision achievable,  in absence of any other error contributions. It
is usually overwhelmed by other systematic contributions. \\

3) {\it Fractional Flux Recovery}: The fractional flux recovery (FFR) quantity is defined as the ratio between the peak brightness in the
astrometric maps (i.e. PRed and MVed maps)
and the self-calibrated maps of the same source. 
It is a useful quantity for comparison between methods
and is related to the thermal noise error. 
It provides an empirical estimate of the residual uncorrelated errors, such as
atmospheric phase fluctuations, which result in
image coherence losses. 
In general,  image coherence losses arise from residual short term phase
fluctuations, hence it is expected to increase with larger duty cycles (in our case the
duty cycle is 300~s). The coherence
losses also increase with the source's angular separation, due to residual long
term phase variations which distort the image.
Nevertheless, neither of these quantities are sensitive to
error processes that cause systematic position offsets, such as those
expected from the spatial structure, direction dependent nature, of
the ionospheric errors. \\

4) {\it Accuracy:} 
The observed quasars were selected from a VLBI catalog with precise positions with accuracies 
$\sim 0.2$ mas  (rfc\_2015b).
Differences of no more than a few mas between the 
catalog positions 
and the relative astrometric measurements presented in
this paper are expected, 
arising from the effects that effectively change the measured
positions. These are:
the use of group delays observable in geodetic analysis compared to the
phase delays in our relative astrometry analysis \citep{porcas_09}, expected
position changes in the observed core at different frequencies
(i.e. the core-shift effect) and differential structure blending
effects between the observing frequencies of the catalog (8.4 GHz) and
our observations (1.6 GHz).

\section{Results}\label{sec:res}

The main goal of these observations was to demonstrate the feasibility
of MV to achieve high precision astrometry 
at low frequencies, along with a
comparative study between MV and PR. 

\subsection{Calibrated Visibility Phases and Astrometric Images: MV vs. PR} \label{sec:results-maps}

\begin{figure}[htbp]
\centering
\includegraphics[width=0.9\textwidth]{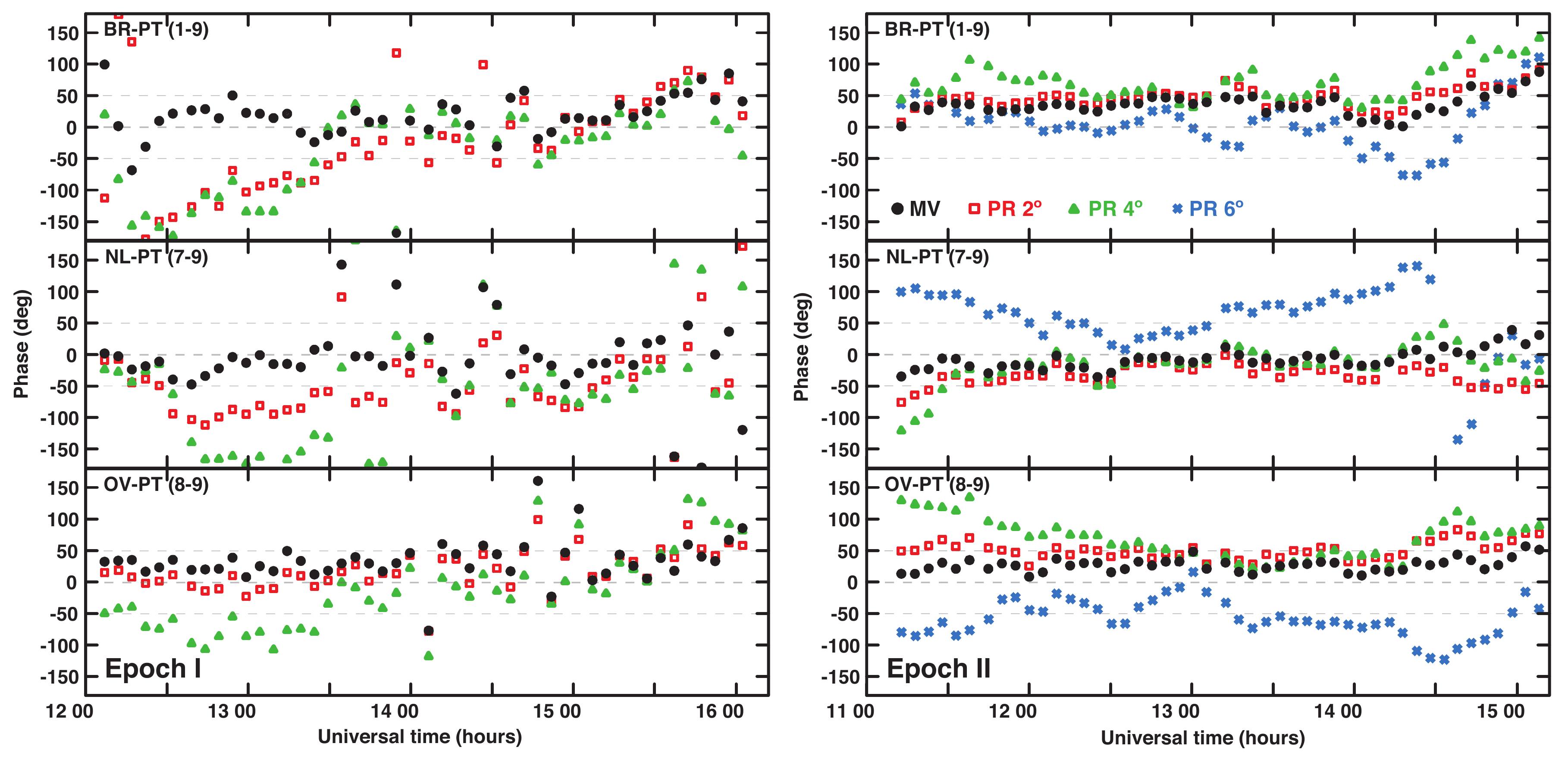}
\caption{Calibrated visibility phases for C4 using MV (black dots)
  with three calibrators, and PR with a calibrator $2^o$ (red
  squares), $4^o$ (green triangles) and $6^o$ (blue crosses, Epoch II only)
  away, on a subgroup of baselines, in Epoch I (left) and Epoch II
  (right).  \label{fig:vis}}
\end{figure}

{\it Visibility Phases:} 
Fig.~\ref{fig:vis} shows a superposition of the residual relative visibility phases
of C4 for a representative subset of baselines, after calibration using
PR (analysis id: PR$_{2^o}$, PR$_{4^o}$ and PR$_{6^o}$) and MV (analysis id: MV$_{QSO}$) 
for the same range of target--calibrator angular separations, from
epochs I (left) and II (right).
The long time-scale trends in PR analysis are easily appreciated:
the deviations from zero are increasingly large for pairs with larger angular
separation and are different in the two epochs, with epoch II being
significantly better . This is indicative of 
systematic residual phase errors, which depend on the weather conditions.
Satellite-based Global Positioning System (GPS) data 
\footnote{The GPS ionospheric data comes from the US Total Electron Content Product Archive of the National Oceanic and Atmospheric Administration (NOAA). https://www.ngdc.noaa.gov/stp/IONO/USTEC/)}
are consistent with epoch II ionospheric conditions being more benign.
The largest disturbances are seen at the beginning of the observations,
which correspond to the sunrise. 
Note that, in epoch II, the phases for the $6^o$  angular separation pair 
are larger (by about a factor of ca. 3) and with opposite sign
compared to those for the  $2^o$ pair and that they correspond to calibrators on opposite sides of the
target. 
This is consistent with the expectations from a wedge-like
ionospheric structure responsible for direction dependent errors as
described in Section~\ref{sec:MVbasis}. 

MV results in the smallest phase residuals (significantly smaller
than PR$_{2^o}$) with almost none of the
signatures for sunrise or systematic trends visible in PR, and are similar for 
both epochs.
All of these are indicative of a superior performance on the
mitigation of ionospheric errors regardless of
the weather conditions. This is crucial for accurate single- and multi-epoch
astrometric analysis, as shown in the next section.

\begin{figure}[htbp]
\centering
\includegraphics[width=0.95\textwidth]{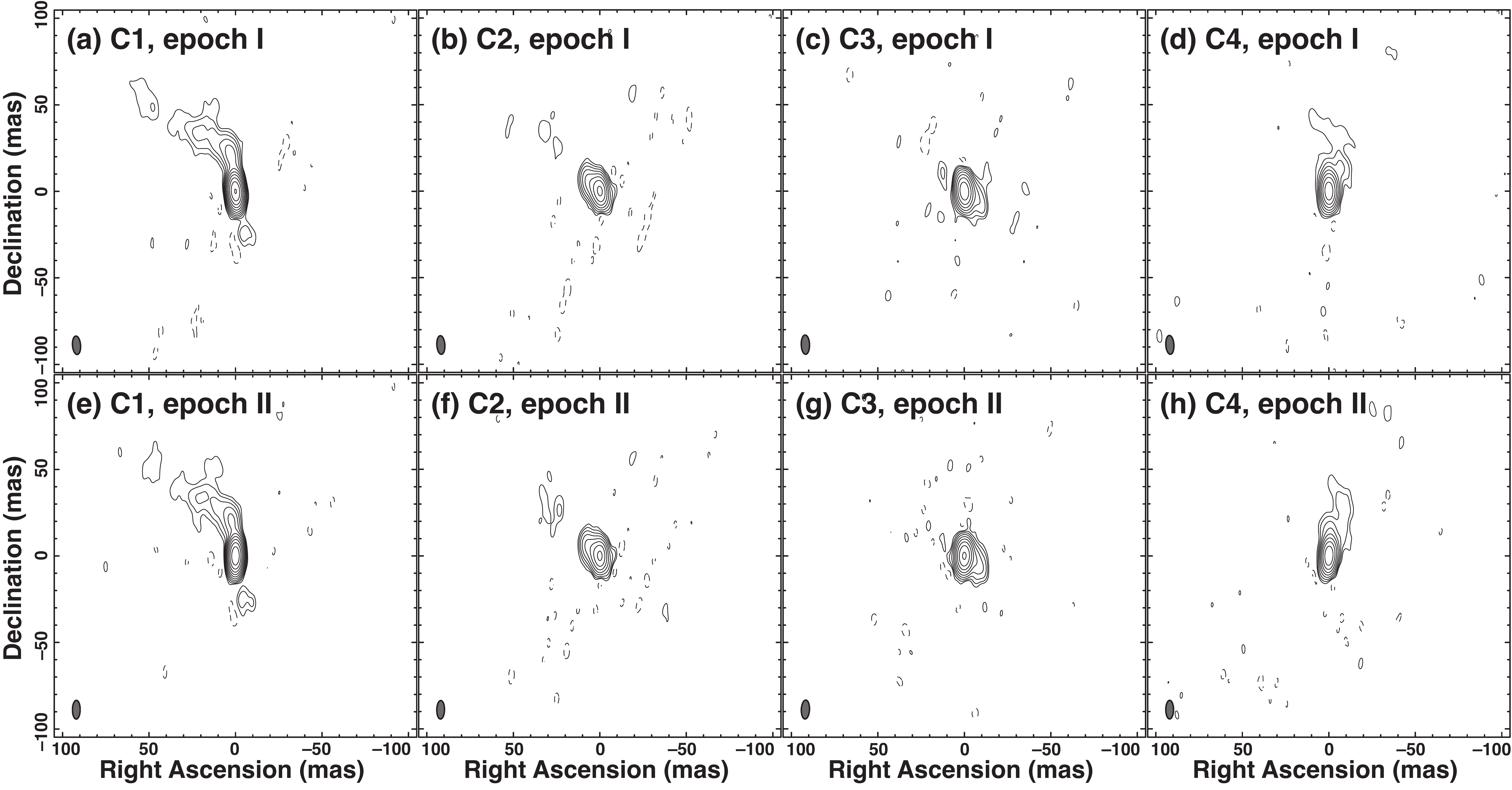}
\caption{Self-calibrated maps of the observed quasars. The peak
  brightness are listed in Table~\ref{table:sources}.
 The lowest intensity contour is the 3-$\sigma$ level and doubling thereafter. Restoring
  beam is $8\times16$~mas with PA=$-10^o$. \label{fig:selfcal.maps}}
\end{figure}

{\it Astrometric images:}  The calibrated visibilities are used to generate the final product of the analysis,
the astrometric images, which convey the astrometric measurements presented in the next section. 
Fig.~\ref{fig:C4.maps} shows the MVed image for the C4 quasar obtained using
the three calibrators together and the three PRed images using 
a single calibrator $\sim \, 2^o,\, 4^o$ and $6^o$ away, 
for the two epochs. 
%
The image degradation arises from remaining short
and long term residual phase variations
in each analysis.
A qualitative comparison suggests that the PRed
images improve with closer angular separations, as
expected; that the MV is similar to PR$_{2^o}$ at both epochs,
with MV slightly better at epoch I, under worse weather conditions.
Fig.~\ref{fig:selfcal.maps} shows the self calibrated maps of the observed quasars.
For a quantitative comparison, Table~\ref{tab:errors} lists the peak
brightness  and rms noise values in the astrometric images along with the
coherence losses (i.e. FFR) estimated with respect to the self calibrated images, for the two epochs of observations.
The corresponding astrometric thermal noise errors, estimated as
described in section~\ref{sec:errors}, are listed in
Table~\ref{table:errors2}. 
Nevertheless, as stated above, these estimates are not sensitive to 
residual systematic errors, which propagate into position shifts
while maintaining the quality of the image.
Those are better addressed by the repeatability errors, presented in the next
section. 

\subsection{Astrometric Repeatability} \label{sec:results-rep}

\begin{figure}[htbp]
\centering
\includegraphics[width=0.8\textwidth]{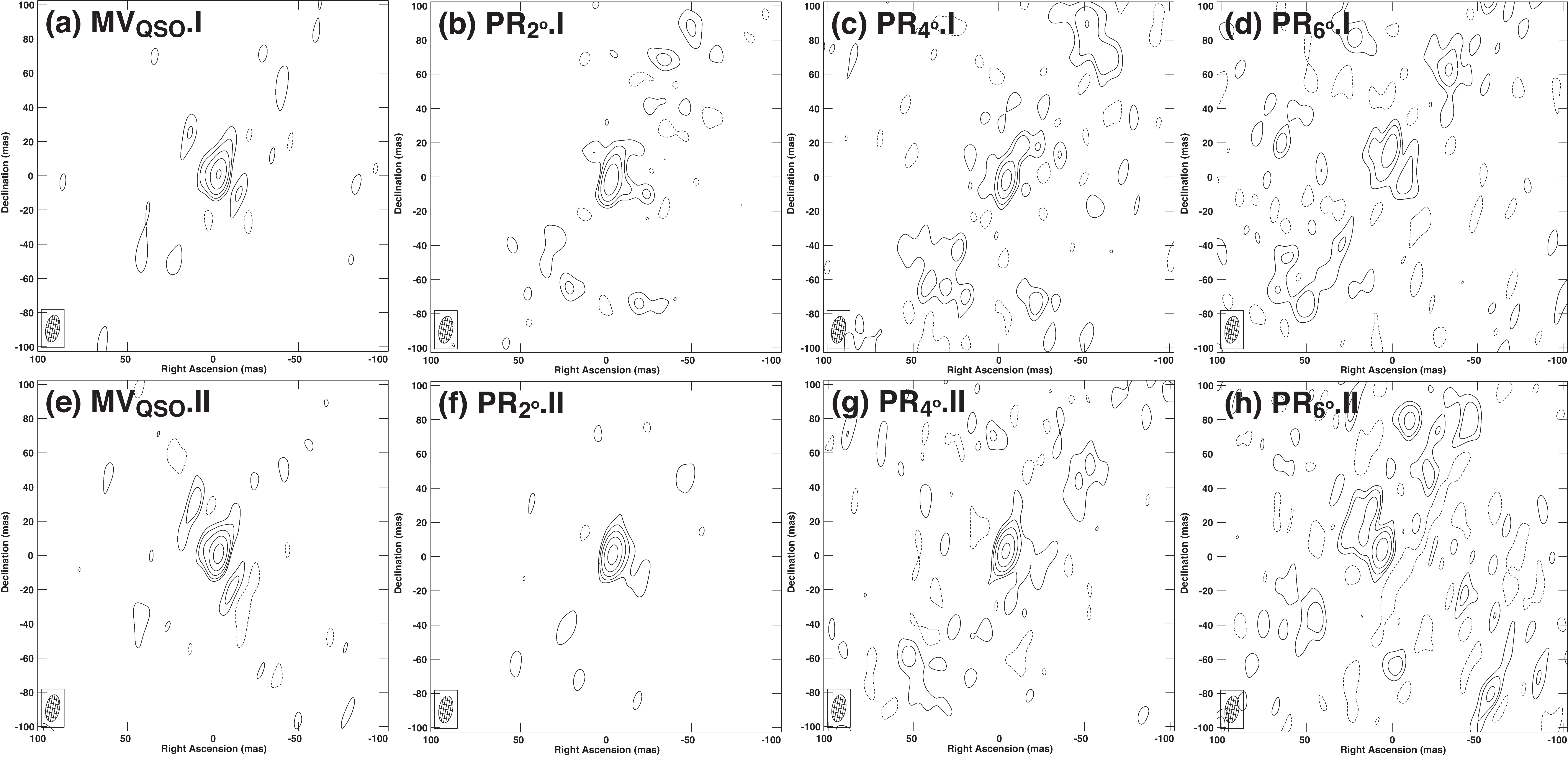}
\caption{Astrometric images of C4 using analyses compared in this paper. From left to right, MV,
  PR$_{2^o}$, PR$_{4^o}$ and PR$_{6^o}$, for epoch I (upper row) and epoch
  II (lower row).  The lowest intensity contour in all images are at 3-$\sigma$
  level of the MV map and doubling thereafter.  The images have
  been restored with the same beam:  $8\times16$~mas with PA=$-10^o$. See
  Table~\ref{tab:errors} for values of peak brightness, rms noise and
  astrometric offsets in the
  images. \label{fig:C4.maps}}
\end{figure}

\begin{figure}[htbp]
\centering
\includegraphics[width=0.9\textwidth]{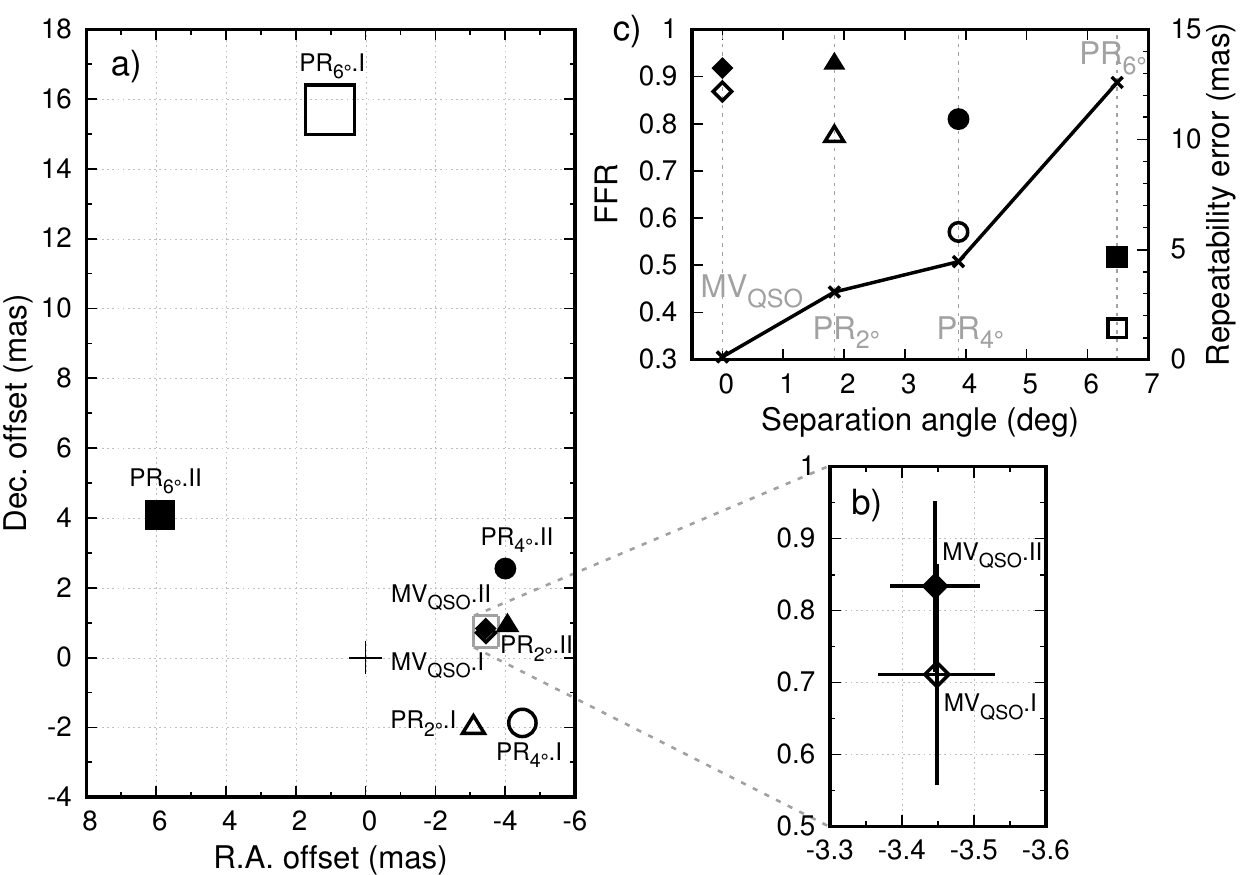}
\caption{ {\it a)} Astrometric offsets in the angular separations
  measured with MV,  PR$_{2^o}$,  PR$_{4^o}$  and
  PR$_{6^o}$ analyses (see Table~\ref{tab:analyses} for description),
  from the observations of quasars at the two
  epochs, with respect to those in the VLBA radio
  fundamental catalog rfc\_2015b. The size of the plotted symbols
  corresponds to the estimated thermal noise error in each case (see
  Table~\ref{table:errors2}). The labels describe the analysis id. and
  epoch of observations. {\it b)}  Zoom for  MV
  astrometric solutions. The error bars are the thermal noise
  errors. Both epochs agree within the error bars.
 {\it c)} Solid line shows the corresponding repeatability astrometric errors versus
 the angular separation between target and calibrator for PR analysis,
 and for an effective $0^o$ separation for MV. Filled and empty
 symbols show the Flux Fractional Recovery quantity versus angular
 separation for MV (diamond), PR$_{2^o}$ (triangle), PR$_{4^o}$ (circle),
 and PR$_{6^o}$ (square) analyses, for epoch I (empty) and epoch II
 (filled). \label{fig:astro}}
\end{figure}

Astrometry is performed directly in the images of quasar C4 shown in Fig.~\ref{fig:C4.maps}
by measuring the offset of the peak of brightness from the center of the map. 
This offset corresponds to the difference between our measurements and the catalog positions used for correlation.
Fig.~\ref{fig:astro}a shows the astrometric measurements, or offsets, of the target quasar C4
using  PR (analysis id. PR$_{2^o}$, PR$_{4^o}$, PR$_{6^o}$) and MV (analysis id.: MV$_{QSO}$)  
at the two epochs of observations. 
Fig.~\ref{fig:astro}b  shows an expansion of the area around the MV
measurements at the two epochs. The astrometric uncertainties in the
plot are 1-$\sigma$ thermal noise error bars ($\sigma_{pos,thermal}$) listed in Table~\ref{table:errors2}.

Note that, in general, while  the PR and MV measurements at a given epoch are expected
to differ, because they are tied to different reference
points, the inter-epoch differences convey information on the source
position changes (if any) undergone in the 1-month time span between
the two epochs. For stationary sources, as it is the case for quasars,
no or negligable position changes are expected.
Therefore one can estimate the repeatability errors 
using the change in the measured offsets at the two epochs for a given technique.
Table~\ref{table:errors2} lists the repeatability errors estimated as described in
Section~\ref{sec:errors} which are an empirical estimate of the
astrometric precision.
It is immediately obvious that 
the repeatability errors are much larger for PR, compared to those for
MV.  Also, that the repeatability errors are larger than the thermal noise errors for PR;
instead they are within the $1\sigma$ thermal noise error bars for MV. \\
Fig.~\ref{fig:astro}c displays the repeatability errors as a
function of source pair angular separations, for PR analysis. 
This linear trend is as expected from PR analysis, 
as closer angular separations provide a better atmospheric 
compensation.
The MV repeatability errors are the
smallest, more than one order of magnitude smaller than those for the
closest pair with PR, and are equivalent to those from a 
very close pair of sources (i.e. close to zero angular separation) in PR analysis. 
It is worth highlighting that instead the MV and PR$_{2^o}$ images
are of similar quality and have similar FFR values. This underlines
the insensitivity of the PRed images to large systematic errors. 
This underlines the superior quality of the calibration of atmospheric errors using
multiple calibrators, compared to that achieved with a single
calibrator with the same range of angular separations, 
and that MV analysis leads to higher precision astrometry. 
This is in agreement with the findings from our previous simulation studies,
where we concluded that using multiple calibrator sources with MV
resulted in one order of magnitude
improvement compared to PR with a single calibrator \citep{sergio,dodson_13}.

Fig.~\ref{fig:astro}a also conveys qualitative information on the astrometric accuracy
of the different calibration techniques 
since the observed quasars have well determined positions
in the rfc\_2015b catalog.
We expect offsets of no more than a few (ca. 1-2) mas to account for
differences between both measurements in all cases (see discussion in Section~\ref{sec:meth}).
Hence the magnitude of each of the astrometric offsets in Fig.~\ref{fig:astro}a  is indicative of the
accuracy of that measurement and the method. PR offsets increase
for larger source separations; 
MV results in the smallest offsets which indicates higher calibration accuracy. 
In fact, the magnitude of the accuracies  is similar to the repeatability errors listed in
Table~\ref{table:errors2}. 

Finally, the distribution of the measurements in Fig.~\ref{fig:astro}a is also indicative of the spatial
structure of the propagation medium being a planar surface.
There is a resemblance between the geometric distribution of the three PR estimates at epoch I
(i.e. PR$_{2^o}$.I,PR$_{4^o}$.I,PR$_{6^o}$.I) and epoch II
(i.e. PR$_{2^o}$.II,PR$_{4^o}$.II,PR$_{6^o}$.II).
Such distributions would arise
from planar spatial structures in the propagation medium above each antenna,
where 
the size and orientation of the triangles 
at each epoch depend on the weather conditions at a given epoch.
The triangle for epoch I is more elongated than for epoch II.
Moreover, both triangles appear rotated with respect to each other around a pivot 
point which is close to the MV measurements,  
which remain practically unchanged (within the thermal noise error bars) at
both epochs, regardless of the weather conditions, as is expected from a quality calibration. 

\subsection{Astrometry on OH-maser source: 
MV vs. in-beam PR}

\begin{figure}[htbp]
\centering
\includegraphics[width=0.9\textwidth]{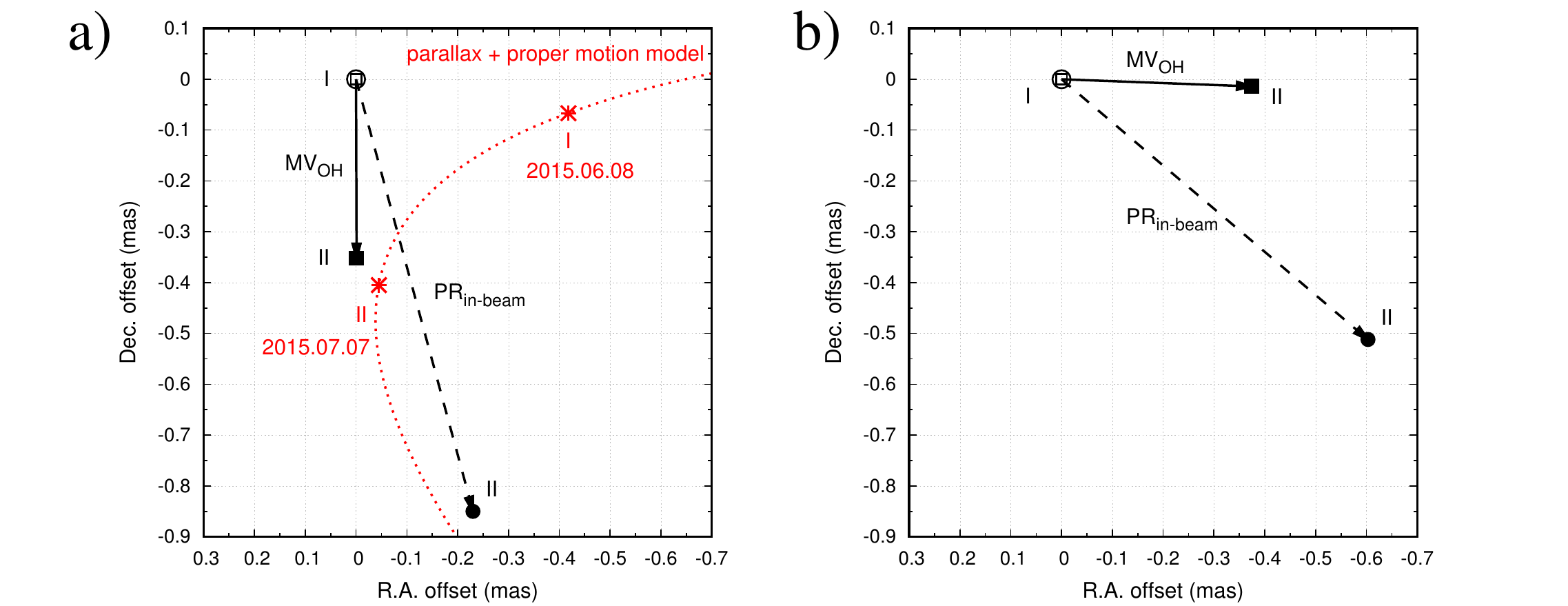}
\caption{{\it a)} Astrometric changes in the position of the line source OH 
 measured with MV and PR$_{in-beam}$ at epoch II, with respect to
 epoch I. The estimated thermal noise errors are 0.5 mas and are not
 displayed.  The labels describe the analysis id. and
  epoch of observations. Also shown is the expected apparent motion of the
  OH-maser source between both epochs, due to the proper motion and
  parallax (see text). {\it b)} Same as {\it left}, after correcting for the apparent
  motion of OH-maser source due to the proper motion and parallax.\label{fig:line}}
\end{figure}

In this section we compare the astrometric results from PR with an
``in-beam'' calibrator
and MV with more distant calibrators, based on the 
analysis using the spectral line OH-maser as the target source
(analysis Ids: PR$_{in-beam}$ and MV$_{OH}$, respectively, in Table~\ref{tab:analyses}).
The PR$_{in-beam}$ analysis uses the simultanteous observations of
C4 and the OH maser target source, 0.4$^o$ away; the MV analysis
uses the alternating observations between the target line source and
the three (continuum) calibrators C1, C2 and C3, which are $\sim\, 2^o$, $4^o$, and $6^o$
away.  
Fig.~\ref{fig:line}a shows the astrometric offsets estimated
at epoch II, with respect to epoch I, for both analyses. 
This accounts for the different reference points in PR and MV and permits a direct comparison of the inter-epoch differences.
The estimated thermal noise errors are
$\sigma_{pos,thermal} \sim \pm 0.5$ mas and $\sim \pm 1$ mas in right ascension and
declination, respectively.
Note that, in this case, the inter-epoch differences
trace the expected motion of the
stellar target,  due to the proper motion and parallax, during the 1
month interval between epochs.
Hence, unlike the case of quasar-only analysis described in the section above, the astrometric
changes between the two epochs are not a direct
measure of the repeatibility errors (and the precision of the method) and an extra step is required to
eliminate the contribution from the motion of the source.  

We used a set of four in-beam phase referencing observations
spanning 1 year to measure the proper motion and parallax
of the OH maser target source \citep{gabor_16}. These include the two epochs 
described in this paper, plus two additional epochs with
a similar observing configuration, except for having longer
duty cycle times ($\sim\, 10$ minutes) which prevented MV analysis. The measured parallax is $\pi
= 2.74 \pm 0.39$ mas and the proper motion is
$\mu(\alpha,\delta)=(-0.17 \pm0.8$ mas yr$^{-1}$, $-7.64\pm0.80$ mas yr$^{-1}$). 
Fig.~\ref{fig:line}b is same as Fig.~\ref{fig:line}a, after removing the contributions from the  proper motion and parallax
between epochs I and II, with the repeatability errors being 
$\sigma^{MV_{OH}}_{pos,rep} = 0.29$ mas for MV$_{OH}$ and $\sigma^{PR_{in-beam}}_{pos,rep} = 0.62$ mas for PR$_{in-beam}$
The repeatability errors for MV$_{OH}$ are half the magnitude for PR$_{in-beam}$,
albeit based on more distant calibrators. In this case all measurements agree within
the thermal noise errors, which are larger for the case of a weak
source. 

\begin{table}[h]
\centering
\begin{tabular}{r r r | c c c | c c}
\hline
\noalign{\smallskip}
Epoch & Analysis Id. & Sep. & Peak & rms &  FFR  &
                                                  \multicolumn{2}{|c}{Astrometric
                                                  Offset} \\
& & (deg) & (mJy/beam) & (mJy/beam) &(\%) & $\Delta_{RA}$ (mas) & $\Delta_{DEC}$ (mas) \\
\hline
I& MV$_{QSO}$ & $\sim 0.$ & 64 & 1  & 87 & -3.45 & 0.71 \\ 
I& PR$_{2^o}$  &  1.8 & 57 & 1 & 77 &  -3.09 & -2.02 \\ 
I & PR$_{4^o}$   & 3.9& 42 & 2 & 57 &  -4.49 & -1.87 \\ 
I & PR$_{6^o}$ & 6.5 & 27 & 2 & 37 & 1.02 & 15.70 \\ 
\hline
II & MV$_{QSO}$  &  $\sim 0.$&  67 & 1 & 92 & -3.45 & 0.83 \\
II & PR$_{2^o}$   &1.8 &  67 & 1 & 93 & -4.06 & 0.90 \\ 
II & PR$_{4^o}$ &  3.9 & 59 & 1 & 81 & -4.01 & 2.55 \\ 
II & PR$_{6^o}$  & 6.5 & 38 & 2 & 52 &  5.89 & 4.09 \\ 
\hline
\end{tabular}
\caption[]{Outcomes from astrometric
  analyses. Column 1 is the epoch of observations. Columns 2 and 3 are the
  analysis id. and source angular separation, respectively. Columns 4, 5 and 6 are the
  peak brightness, rms noise and Fractional Flux Recovery values measured from the astrometric images
  in Fig.~\ref{fig:C4.maps}, respectively.
 Columns 7 and 8 are the astrometric offsets of the peak of brightness in the astrometric images shown 
  in Fig.~\ref{fig:C4.maps} from the centre of the map; these are shown  
in Fig.~\ref{fig:astro}a). \label{tab:errors}}
\end{table}

\begin{table}[h]
\centering
\begin{tabular}{l| c c |c}
\hline
\noalign{\smallskip}
Analysis Id. &        \multicolumn{2}{|c|}{$\sigma_{pos,thermal}$
               (mas)} &  $\sigma_{pos,rep}$ (mas)  \\
            &  Epoch  I & Epoch II &  $\Delta_{I-II}^\prime$ \\
\noalign{\smallskip}
\hline
MV$_{QSO}$ & 0.17 & 0.14 & 0.1 \\
PR$_{2^o}$  &  0.19 & 0.11 & 2.7  \\
PR$_{4^o}$   & 0.42 & 0.22 & 3.9 \\
PR$_{6^o}$ & 0.75 & 0.44 & 11.2  \\
\noalign{\smallskip}
\hline
\end{tabular}
\caption[]{\label{table:errors2}Empirically estimated astrometric errors for measurements presented in this paper.
Column 1 is the calibration strategy (see Table~\ref{tab:analyses} for
description). Columns 2 and 3 are the thermal noise errors for epochs
I and II, respectively. They have been calculated using the values
listed in Table~\ref{tab:errors} with a beam of $8\times16$~mas PA=-10$^o$. 
Column 4 lists the repeatability errors, calculated from the 
astrometric offsets in Table~\ref{tab:errors} and corrected for bias
for two epochs, 
with $\Delta_{I-II}^\prime$ =$\Delta_{I-II} / \sqrt {2} \ast \sqrt{2 /\pi}$.
See Section~\ref{sec:errors} for description of error analysis.
} 
\end{table}

\section{Discussion and Conclusions}
\label{sec:disc}

\noindent
{\it Demonstration of MultiView high precision astrometry at low frequencies}\\
The ionospheric propagation effects are the main limitation to
routinely achieving  high precision astrometry at frequencies $ca. <8$~GHz,
using state-of-the-art phase referencing methods developed for higher frequencies. 
This is due to the distinct direction dependent signature,
which limits high precision measurements to cases
when there is a suitable very close calibrator ca. arcmins away.
Our combination of multiple calibrators around the
target results in a significant reduction of the systematic
astrometric errors, from the mitigation of the spatial structure effects,
by using a 2-D linear (spatial) interpolation to estimate the calibration along the direction of the target source.\\
We have presented an empirical demonstration of the superior
mitigation of MV along with a comparative study with phase referencing
analysis using a single source at a range of angular separations, including ``in-beam'' phase referencing.
All the analysis have been carried out using the same VLBA observations at 1.6 GHz. 
We have used the repeatability between two
epochs of observations to provide an empirical estimate 
of the systematic astrometric errors, which are otherwise very difficult to quantify.

We achieve high precision MV astrometry  of ca. $100~\mu$as
in a single epoch of observations of quasars with calibrators 
at $2^o$ and larger angular separations, effectively reaching the thermal noise limit of the observations.
This corresponds to more than an order of magnitude
improvement with respect to the precision achieved using PR with a
single calibrator $2^o$ away, which is ca. $3\,$ mas, due to the residual systematic errors in the analysis.
This underlines the importance of correcting for the spatial structure
of the ionospheric residuals. 
The comparative improvement can be interpreted as MV compensation being equivalent to
that from  PR with a single calibrator ca. 10 times closer, in this
case ca. $0.2^o$ away, assuming a linear  dependence between astrometric precision and target--calibrator angular separation.  
Also, our results indicate that there is a common limiting factor for accuracy and precision, 
namely the residual ionospheric propagation errors, and both are improved by a quality calibration as provided by MV.

We have also demonstrated the performance in the weak source case
using the observations of the OH-maser line source, and compared MV
using $2^o$, $4^o$ and $6^o$ angular separations 
with \
`in beam' PR using a calibrator $0.4^o$ away. The repeatability
errors are larger, as expected from lower SNR, but interestingly keep the same
corresponding relative astrometric signature found in the analysis of
the quasars. That is MV$_{OH}$ calibration is a factor of two better,
with respect to PR$_{in-beam}$ using a calibrator five times closer to the target.
Therefore we have demonstrated the benefits of using multiple
calibrators, in our case with an improvement of more than one order of
magnitude in astrometric precision, reaching the thermal limit of
the observations, of $\sim 100 \mu$as.  
In general we expect MV to be relevant for observations in the frequency regime where 
the ionospheric effects continue to be the dominant source of errors, that is, in observations up to ca. 8 GHz. 
We conclude that greater improvements are expected from increased
sensitivity, and faster duty cycles, with maximum benefits from
simultaneous observations and closer source distribution, to minimize
the non-linear deviations of the actual ionospheric spatial structure
above each antenna from a planar surface. 

\noindent
{\it MV in the context of SKA and multi-beam instruments} \\
Precise astrometric capability is of great importance in the SKA
era. It is a SKA goal to achieve $10 \, \mu$as astrometric
accuracy at a single epoch of observations at $\sim$1.4~GHz
\citep{ska_vlbi}. 
The high sensitivity and long baselines of SKA VLBI
observations  will result in a much
reduced thermal noise level and high spatial resolution. Therefore this goal
is achievable as long as a sufficiently accurate
ionospheric phase calibration strategy is in place. 
For a single calibrator source and PR techniques, the required
angular separation to the target would be ca. 1 arcmin \citep{ska_vlbi}.
This puts  a very tight limit
on the number of available calibrator sources, even at the
SKA sensitivities \citep{ska_135}. 
This constraint on the angular separation can be significantly relaxed,
by using multiple calibrator sources and MV techniques,
as suggested by the demonstration presented in this paper and our
previous simulation studies.  Additionally, the multi-beam capability
of SKA will allow for simultaneous observations of all sources and
therefore eliminate the errors arising from short term phase
fluctuations, which result in a reduction of the coherence losses
(i.e. characterized by FFR quantity) and the thermal noise errors,
while improving the overall performance. This applies to other
instruments with multibeam systems such as ASKAP and Westerbork
Synthesis Radio Telescope with Apertif, among others.  Our
demonstration includes three calibrators; note that the more
calibrators the better, as this will allow the most accurate
reconstruction of the atmospheric effects.  Therefore we expect MV can
deliver the goal of $10\,\mu$as astrometry for many targets with SKA.

{\it Other relevant scientific applications of MV} \\
Next we consider other scientific applications. 

\paragraph{Near Field Cosmology:}
This paper used astrometric observations of a group of quasars in the role of
targets and calibrators and an OH maser source in the ground
state at 1.6 GHz. Scientific applications using such a group
of sources applied to studies with the SKA of the nearby universe, including the
Milky Way galaxy and the Local Group of galaxies 
are described in \citet{imai_ska_16}.
Although that case-study report discussed the
scientific applications conservatively based on only `in-beam' PR
astrometry, our results indicate that MV would provide further benefits.

\paragraph{Pulsars at 1.6 GHz:}
Our  empirically estimated MV astrometric  accuracy of ca. $100\, \mu$as
at 1.6 GHz with VLBA observations, using calibrators more than ca. $2^o$ away
is at the state-of-the-art, only comparable with 
`in-beam' phase referencing observations with a  calibrator
ca. 10s of arcminutes away  \citep{deller_13, deller_16}.
This improvement is expected to continue to apply at all angular
scales. Hence, using three calibrators within the SKA-Mid antenna beams and MV will result in
a further increase
by one order of magnitude of the astrometric precision, extrapolating from our comparative study. 
In general, allowing for larger angular separations makes it possible 
to select good calibrator sources, 
which are fundamental for
multi-epoch studies.  With the higher probabilities of finding
suitable calibrators 
the general applicability is also highly increased.
In some cases, such as pulsar observations in the galactic plane, it
might even be desirable to use calibrators out of the plane to reduce the
effect of scattering. This would be possible using MV.

\paragraph{Methanol masers at 6.7 GHz:}
High precision astrometry observations of methanol masers at  6.7 GHz
holds the prospect to contribute to the successful program for 3-D
mapping of our Galaxy, as a complement to the precise 
water maser measurements \citep{reid_14}. However the advanced PR strategies used at 22 GHz, in the
tropospheric dominated regime, fail to provide
high precision astrometry at 6.7 GHz, as this is in the ionospheric dominated
regime. Therefore MV with fast source switching between
sources, or simultaneous observations if possible, 
provides a strategy for superior calibration of
the tropospheric and ionospheric errors resulting in precise
astrometry.   \\
\noindent
{\bf Acknowledgements}

\noindent
The VLBA is operated by National Radio Astronomy Observatory and is a facility of the National Science Foundation operated under cooperative agreement with Associated Universities Inc..  
GO and HI have been supported by the JSPS Bilateral Collaboration
Program and KAKENHI programs 25610043 and 16H02167. MR, RD, GO and HI
acknowledge support from DFAT grant AJF-124.
%
This research has made use of the SIMBAD database,
operated at CDS, Strasbourg, France. 



\begin{thebibliography}{26}
\expandafter\ifx\csname natexlab\endcsname\relax\def\natexlab#1{#1}\fi

\bibitem[{{Alef}(1988)}]{alef_88}
{Alef}, W. 1988, in IAU Symposium, Vol. 129, The Impact of VLBI on Astrophysics
  and Geophysics, ed. M.~J. {Reid} \& J.~M. {Moran}, 523

\bibitem[{{Deller} {et~al.}(2013){Deller}, {Boyles}, {Lorimer}, {Kaspi},
  {McLaughlin}, {Ransom}, {Stairs}, \& {Stovall}}]{deller_13}
{Deller}, A.~T., {Boyles}, J., {Lorimer}, D.~R., {et~al.} 2013, \apj, 770, 145

\bibitem[{{Deller} {et~al.}(2016){Deller}, {Vigeland}, {Kaplan}, {Goss},
  {Brisken}, {Chatterjee}, {Cordes}, {Janssen}, {Lazio}, {Petrov}, {Stappers},
  \& {Lyne}}]{deller_16}
{Deller}, A.~T., {Vigeland}, S.~J., {Kaplan}, D.~L., {et~al.} 2016, \apj, 828,
  8

\bibitem[{{Dodson} {et~al.}(2013){Dodson}, {Rioja}, {Asaki}, {Imai}, {Hong}, \&
  {Shen}}]{dodson_13}
{Dodson}, R., {Rioja}, M., {Asaki}, Y., {et~al.} 2013, AJ, 145, 147

\bibitem[{{Dodson} {et~al.}(2016){Dodson}, {Rioja}, {Molina}, \&
  {G\'omez}}]{dodson_16}
{Dodson}, R., {Rioja}, M., {Molina}, S., \& {G\'omez}, J. 2016, Submitted

\bibitem[{{Doi} {et~al.}(2006){Doi}, {Fujisawa}, {Habe}, {Honma}, {Kawaguchi},
  {Kobayashi}, {Murata}, {Omodaka}, {Sudou}, \& {Takaba}}]{doi_06}
{Doi}, A., {Fujisawa}, K., {Habe}, A., {et~al.} 2006, \pasj, 58, 777

\bibitem[{{Fomalont} {et~al.}(1999){Fomalont}, {Goss}, {Beasley}, \&
  {Chatterjee}}]{fomalont_99}
{Fomalont}, E.~B., {Goss}, W.~M., {Beasley}, A.~J., \& {Chatterjee}, S. 1999,
  \aj, 117, 3025

\bibitem[{{Fomalont} \& {Kopeikin}(2002)}]{fomalont_02}
{Fomalont}, E.~B., \& {Kopeikin}, S. 2002, in Proceedings of the 6th EVN
  Symposium, ed. E.~{Ros}, R.~W. {Porcas}, A.~P. {Lobanov}, \& J.~A. {Zensus},
  53

\bibitem[{{Godfrey} {et~al.}(2011){Godfrey}, {Bignall}, \& Tingay}]{ska_135}
{Godfrey}, L., {Bignall}, H., \& Tingay, S. 2011, {Very High Angular Resolution
  Science with the SKA}, Tech. rep., Curtain University, Australia

\bibitem[{{Greisen}(2003)}]{aips}
{Greisen}, E.~W. 2003, Information Handling in Astronomy - Historical Vistas,
  285, 109

\bibitem[{{Honma} {et~al.}(2008){Honma}, {Tamura}, \& {Reid}}]{honma_08_trop}
{Honma}, M., {Tamura}, Y., \& {Reid}, M.~J. 2008, \pasj, 60, 951

\bibitem[{{Imai} {et~al.}(2016){Imai}, {Burns}, {Yamada}, {Goda}, {Yano},
  {Orosz}, {Niinuma}, \& {Bekki}}]{imai_ska_16}
{Imai}, H., {Burns}, R.~A., {Yamada}, Y., {et~al.} 2016, ArXiv e-prints

\bibitem[{{Jimenez-Monferrer} {et~al.}(2010){Jimenez-Monferrer}, {Rioja},
  {Dodson}, {Smirnov}, \& {Guirado}}]{sergio}
{Jimenez-Monferrer}, S., {Rioja}, M.~J., {Dodson}, R., {Smirnov}, O., \&
  {Guirado}, J.~C. 2010, in 10th European VLBI Network Symposium and EVN Users
  Meeting: VLBI and the New Generation of Radio Arrays, Manchester, Proceedings
  of Science, Vol. 125, 84

\bibitem[{{Orosz} {et~al.}(2016){Orosz}, {Imai}, {Dodson}, {Rioja}, S., Burns
  R.~Engels, Etoka, Nakagawa, Nakanishi, Asaki, Goldman, Nanni, Marigo, \&
  Tafoya}]{gabor_16}
{Orosz}, G., {Imai}, H., {Dodson}, R., {et~al.} 2016, Submitted

\bibitem[{{Paragi} {et~al.}(2014){Paragi}, {Godfrey}, {Reynolds}, {Rioja},
  {Deller}, {Zhang}, {Gurvits}, {Bietenholz}, {Szomoru}, {Bignall}, {Boven},
  {Charlot}, {Dodson}, {Frey}, {Garrett}, {Imai}, {Lobanov}, {Reid}, {Ros},
  {van Langevelde}, {Zensus}, {Zheng}, {Alberdi}, {Agudo}, {An}, {Argo},
  {Beswick}, {Biggs}, {Brunthaler}, {Campbell}, {Cimo}, {Colomer}, {Corbel},
  {Conway}, {Cseh}, {Deane}, {Falcke}, {Gabanyi}, {Gawronski}, {Gaylard},
  {Giovannini}, {Giroletti}, {Goddi}, {Goedhart}, {Gomez}, {Gunn}, {Jung},
  {Kharb}, {Klockner}, {Kording}, {Kovalev}, {Kunert-Bajraszewska},
  {Lindqvist}, {Lister}, {Mantovani}, {Marti-Vidal}, {Mezcua}, {McKean},
  {Middelberg}, {Miller-Jones}, {Moldon}, {Muxlow}, {O'Brien},
  {P{\'e}rez-Torres}, {Pogrebenko}, {Quick}, {Rushton}, {Schilizzi}, {Smirnov},
  {Sohn}, {Surcis}, {Taylor}, {Tingay}, {Tudose}, {van der Horst}, {van
  Leeuwen}, {Venturi}, {Vermeulen}, {Vlemmings}, {de Witt}, {Wucknitz}, \&
  {Yang}}]{ska_vlbi}
{Paragi}, Z., {Godfrey}, L., {Reynolds}, C., {et~al.} 2014, ArXiv e-prints

\bibitem[{{Porcas}(2009)}]{porcas_09}
{Porcas}, R.~W. 2009, \aap, 505, L1

\bibitem[{{Porcas} \& {Rioja}(2002)}]{porcas_02}
{Porcas}, R.~W., \& {Rioja}, M.~J. 2002, in Proceedings of the 6th EVN
  Symposium, ed. E.~{Ros}, R.~W. {Porcas}, A.~P. {Lobanov}, \& J.~A. {Zensus},
  65

\bibitem[{{Reid} \& {Brunthaler}(2004)}]{reid_04}
{Reid}, M.~J., \& {Brunthaler}, A. 2004, \apj, 616, 872

\bibitem[{{Reid} \& {Honma}(2014)}]{reid_micro}
{Reid}, M.~J., \& {Honma}, M. 2014, ARAA, 52, 339

\bibitem[{{Reid} {et~al.}(2014){Reid}, {Menten}, {Brunthaler}, {Zheng}, {Dame},
  {Xu}, {Wu}, {Zhang}, {Sanna}, {Sato}, {Hachisuka}, {Choi}, {Immer},
  {Moscadelli}, {Rygl}, \& {Bartkiewicz}}]{reid_14}
{Reid}, M.~J., {Menten}, K.~M., {Brunthaler}, A., {et~al.} 2014, \apj, 783, 130

\bibitem[{{Rioja} \& {Dodson}(2011)}]{rioja_11a}
{Rioja}, M., \& {Dodson}, R. 2011, AJ, 141, 114

\bibitem[{{Rioja} {et~al.}(2009){Rioja}, {Dodson}, {Porcas}, {Ferris},
  {Reynolds}, {Sasao}, \& {Schilizzi}}]{rioja_09}
{Rioja}, M., {Dodson}, R., {Porcas}, R.~W., {et~al.} 2009, in 8th International
  e-VLBI Workshop, 14

\bibitem[{{Rioja} {et~al.}(2002){Rioja}, {Porcas}, {Desmurs}, {Alef},
  {Gurvits}, \& {Schilizzi}}]{rioja_02}
{Rioja}, M.~J., {Porcas}, R.~W., {Desmurs}, J.-F., {et~al.} 2002, in
  Proceedings of the 6th EVN Symposium, ed. E.~{Ros}, R.~W. {Porcas}, A.~P.
  {Lobanov}, \& J.~A. {Zensus}, 57

\bibitem[{{Rioja} {et~al.}(1997){Rioja}, {Stevens}, {Gurvits}, {Alef},
  {Schilizzi}, {Sasao}, \& {Asaki}}]{rioja_97}
{Rioja}, M.~J., {Stevens}, E., {Gurvits}, L., {et~al.} 1997, Vistas in
  Astronomy, 41, 213

\bibitem[{{Shepherd} {et~al.}(1994){Shepherd}, {Pearson}, \& {Taylor}}]{difmap}
{Shepherd}, M.~C., {Pearson}, T.~J., \& {Taylor}, G.~B. 1994, in Bulletin of
  the American Astronomical Society, Vol.~26, Bulletin of the American
  Astronomical Society, 987--989

\bibitem[{Wrobel {et~al.}(2000)Wrobel, Walker, Benson, \& A..}]{vlba_24}
Wrobel, J., Walker, R., Benson, J., \& A.., B. 2000, {VLBA Scientific
  Memorandum n. 24: Strategies for Phase Referencing with the VLBA}, Tech.
  rep., NRAO

\end{thebibliography}

\end{document}